# Deep learning acceleration of iterative model-based light fluence correction for photoacoustic tomography

Zhaoyong Liang[a,b,c], Shuangyang Zhang[a,b,c], Zhichao Liang[a,b,c], Zhongxin Mo[a,b,c], Xiaoming Zhang[a,b,c], Yutian Zhong[a,b,c], Wufan Chen[a,b,c], Li Qi[a,b,c],*

[a] School of Biomedical Engineering, Southern Medical University, 1023 Shatai Rd., Baiyun District, Guangzhou, Guangdong, 510515, China

[b] Guangdong Provincial Key Laboratory of Medical Image Processing, Southern Medical University, 1023 Shatai Rd., Baiyun District, Guangzhou, Guangdong, 510515, China

[c] Guangdong Province Engineering Laboratory for Medical Imaging and Diagnostic Technology, Southern Medical University, 1023 Shatai Rd., Baiyun District, Guangzhou, Guangdong, 510515, China

* Corresponding author: qili@smu.edu.cn (L. Qi)

**Abstract**

Photoacoustic tomography (PAT) is a promising imaging technique that can visualize the distribution of chromophores within biological tissue. However, the accuracy of PAT imaging is compromised by light fluence (LF), which hinders the quantification of light absorbers. Currently, model-based iterative methods are used for LF correction, but they require significant computational resources due to repeated LF estimation based on differential light transport models. To improve LF correction efficiency, we propose to use Fourier neural operator (FNO), a neural network specially designed for solving differential equations, to learn the forward projection of light transport in PAT. Trained using paired finite-element-based LF simulation data, our FNO model replaces the traditional computational heavy LF estimator during iterative correction, such that the correction procedure is significantly accelerated. Simulation and experimental results demonstrate that our method achieves comparable LF correction quality to traditional iterative methods while reducing the correction time by over 30 times.

1. Introduction

Photoacoustic tomography (PAT) is a non-invasive imaging modality that combines the benefits of optical and ultrasound imaging. PAT utilizes the transient thermoelastic expansion of biological tissues, excited by light, to generate acoustic waves that are subsequently detected by an array of sensors [1]-[4]. A cross-sectional image representing the initial absorbed light energy is obtained using the detected signal. Using multi-wavelength light excitation, multispectral PAT is able to effectively distinguish and quantitatively analyze different biological tissues given their optical absorption at different excitation wavelengths. Profiting from high imaging speed and centimeter-scale imaging depth, Multispectral PAT has been widely applied in various preclinical studies [5][6] and clinical trials [7][8].

PAT image represents the initial PA pressure after laser excitation, which is proportional to the absorbed energy per unit volume of tissue. The absorbed energy density can be approximated by the product of the light fluence (LF) arriving at a voxel and the absorption coefficient ($\mu_a$) of the enclosed absorber [9]. Removing LF, the quantitative $\mu_a$ map which represents intrinsic optical features of specific absorbers can be obtained [10]. However, since measuring the actual LF distribution of the imaging object is challenging, light transport models such as radiation transfer equation [11][12], diffusion equation [13][14], and numerical models based on Monte Carlo method [15][16] are employed to simulate the deposited LF.

In an ideal case, if the estimated LF distribution is accurate, the quantitative $\mu_a$ map can be calculated easily by dividing initial pressure by LF. In practice, however, there is always error and variation during not only LF simulation but initial pressure reconstruction as well. This leads to inaccuracy in the simulation and correction of LF. To solve this problem, model-based iterative LF correction algorithm is proposed [17]-[21]. This method does not require the analytic form of the model, instead, it continually updates the unknown parameters through iterative optimization until the output of the solver matches the measured data [9]. For example, Yuan et al. [17] used an iterative nonlinear algorithm based on the diffusion equation and combined Tikonov regularization with space-based regularization to restore the optical properties of PAT images. Cox et al. [18][19] accelerated the calculation of the functional gradient vector with an adjoint model to achieve greater computational efficiency, but resulting in more iterations to converge. Liu et al. [20] applied an iterative algorithm based on 3D Monte Carlo simulation of light transport to achieve 3D light correction in PAT, but it was limited to simple numerical simulation and phantom experiments. In order to improve the correction accuracy and simplify the calculation process, Zhang et al. [21] proposed a non-segmentation pixel-wise correction method by using a finite-element-based model to iteratively optimize both the $\mu_a$ map and LF distribution of target tissues. This method has shown excellent correction effects in simulation, phantom, and animal experiments.

Although these model-based iterative LF correction methods have been well established, they all face a major problem: heavy computational burden. The reason causing such a problem is that during iteration, a numerical LF solver is repeatedly used to update the LF distribution within the object. Whether the solver is based on finite element method (FEM) or Monte Carlo method, the solution of an accurate LF map requires high computational cost. For the acceleration of LF solvers, graphics processing unit (GPU) is used without reducing the total computation amount [22][23]. To reduce computation time, researchers have also proposed adding tissue segmentation results as prior knowledge to assist correction [24]-[26], yet its estimation accuracy is compromised since a uniform distribution of optical coefficient is assumed within each segmented tissue region.

Except for the above model-based correction methods, deep-learning-based (DL-based) approaches have been introduced to LF correction of PAT imaging. Currently, the application of DL is mainly focused on directly reconstructing absorption coefficient maps from original PAT images [27]-[29]. For example, Chen et al. [28] proposed a DL-based quantitative photoacoustic imaging method based on U-Net, which obtained the absorption coefficient map directly from the reconstructed initial pressure maps. The effectiveness of their method is proved by simulation experiments based on Monte Carlo LF simulation. By comparing various network structures, Madasamy et al. [29] found that DL-based method could compensate for the nonlinear light fluence distribution more effectively and more efficiently as well. However, due to the need to learn an unknown mapping, DL-based methods usually require a large amount of data and a large-size network for training and testing. Their performance is still inferior to traditional iterative correction approaches.

For conventional LF correction techniques based on the diffusion equation, the computational bottleneck is centered on solving the light-transport partial differential equation (PDE). Recently, efficient solvers for PDE have been explored using DL-based techniques, which have shown great promise in speed and accuracy. These DL-based PDE solvers include finite-dimensional operators [30]-[32], neural finite element models [33]-[35], and neural operators [36]-[39]. Compared to their opponents, neural operators overcome the mesh-dependent nature of finite-dimensional approaches by generating a single set of network parameters used with different discretization. They only need to be trained once, and they do not require knowledge of the underlying PDE and are thus suitable for problems where accurate PDEs are difficult to obtain. The recently reported Fourier neural operator (FNO) shares all of the mentioned characteristics of neural operators. By parameterizing the integral kernel in Fourier space, it is able to further reduce the cost of evaluating the integration operator [39].

Considering its potential advantage in LF estimation, neural operators may be used to replace the LF estimator in model-based iterative correction, such that the heavy computational cost of repeated LF calculation can be reduced. Based on this concept, herein we propose a FNO-based accelerated iterative light fluence correction method to speed up iterative LF correction in PAT imaging. Our method is based on the model-based iterative LF correction algorithm, but we couple a trained FNO network model into the iteration process as a forward LF estimator and update both the $\mu_a$ map and LF distribution through alternating optimization. Due to its invariance to discretization and minimal network size, FNO is chosen to learn the light transport model, which enables precise LF estimation result with fewer computing resource. With our FNO-based acceleration technique, the iterative LF correction procedure achieves an over 30-fold increase in correction speed compared to traditional methods, as verified by simulation and small animal imaging experiments. Despite this significant improvement in processing time, the imaging quality obtained by the proposed method is still comparable to that obtained using traditional iterative correction methods.

## 2. Methods

### 2.1 The optical forward process in PAT imaging

During the optical forward process in PAT imaging, the initial pressure distribution $p(r)$ at a point $r$ within a given biological tissue can be expressed by the following formula:

$$p(r) = \Gamma \mu_a(r) \phi(\mu_a(r), \mu_s(r), g(r)), \quad (1)$$

where $\Gamma$ denotes the Gruneisen coefficient, which represents the PA efficiency, measuring the transformation efficiency from thermal energy to pressure. $\mu_a(r)$ and $\mu_s(r)$ denote local absorption and reduced scattering coefficients respectively. $\phi$ represents the deposited light fluence, and $g(r)$ represents the anisotropic scattering factor at the point $r$. Let us assume that the PAT image is accurately reconstructed and the effect of structural distortion is ignored [41][42]. Also, in biological soft tissues, the Gruneisen coefficient has a small change, so it is assumed to be constant [25]. In this case, the photoacoustic pressure $p'$ in the reconstructed image can be expressed as the product of the absorption

coefficient $\mu_a$ and the LF distribution $\phi$:

$$p'(r) = \phi(\mu_a(r), \mu_s'(r)) \cdot \mu_a(r), \tag{2}$$

where $\mu_s'(r)$ is the reduced scattering coefficient, calculated by $\mu_s'(r) = \mu_s(r)(1 - g(r))$.

**2.2 Model-based iterative light fluence correction**

In this work, we choose diffusion equation (DE) as our light transport model. DE is the first-order spherical harmonic expansion approximation of the radiative transfer equation. It is given in the frequency domain as:

$$-\nabla \cdot D(r)\nabla\phi(r,\omega) + \left(\mu_a(r) + \frac{i\omega}{c}\right)\phi(r,\omega) = q(r), \tag{3}$$

where $k = c/(3(\mu_a + \mu_s'))$, $c$ represents speed of light. $q$ represents the light source, which can be viewed as constant in both time and space in PAT. It can be seen that the LF distribution $\phi$ is co-determined by $\mu_a$ and $\mu_s'$. Therefore, the solution of (3) can be obtained by finite-element methods given the distribution of $\mu_a$ and $\mu_s'$.

In biological tissue, since the variation of tissue scattering coefficient is small, $\mu_s'$ is often assumed known [10][25][43][44]. Therefore, the forward model (2) can be simplified to:

$$p' = \phi(\mu_a) \cdot \mu_a, \tag{4}$$

where $p'$ and $\mu_a$ are in vector representation, $\phi$ is LF in sparse matrix representation. Accordingly, the solution of $\mu_a$ can be formulated as a least square optimization problem:

$$\mu_a^s = arg\ min\|p' - \phi(\mu_a) \cdot \mu_a\|^2. \tag{5}$$

This problem can be solved by the alternating iterative optimization method, as shown in Fig. 1(a). To begin with, the $\mu_a$ map is initialized as 0, based on which an initial LF map is obtained accordingly by using a FEM-based DE solver. In each iteration, a new LF matrix $\phi^k$ is first solved by the LF solver given the last updated absorption map $\mu_a^{k-1}$. With $\phi^k$ fixed, the optimization problem is converted into a linear model and a new $\mu_a^k$ map is obtained by using the gradient descent algorithm. Then the $\mu_a^k$ map is used as the input for the next iteration. The iterative procedure continues until the residual error is smaller than a predefined value or the number of iterations exceeds a preset value, and the final $\mu_a$ map represents the corrected PAT image.

The above iterative LF correction method combines light transport model with iterative optimization to reconstruct the optical coefficient map of the target. By continuously updating the $\mu_a$ map and LF map, the difference between the measurement data and model-generated data is minimized, and a satisfactory $\mu_a$ image is obtained.

However, the above iterative correction requires repeated estimation of the LF distribution $\phi^k$, i.e. repeated utilization of the numerical solver for light transport modeling. Since traditional LF estimators are based on either the Monte Carlo method or FEM methods, the computational cost for these methods is relatively high, leading to a long correction time.

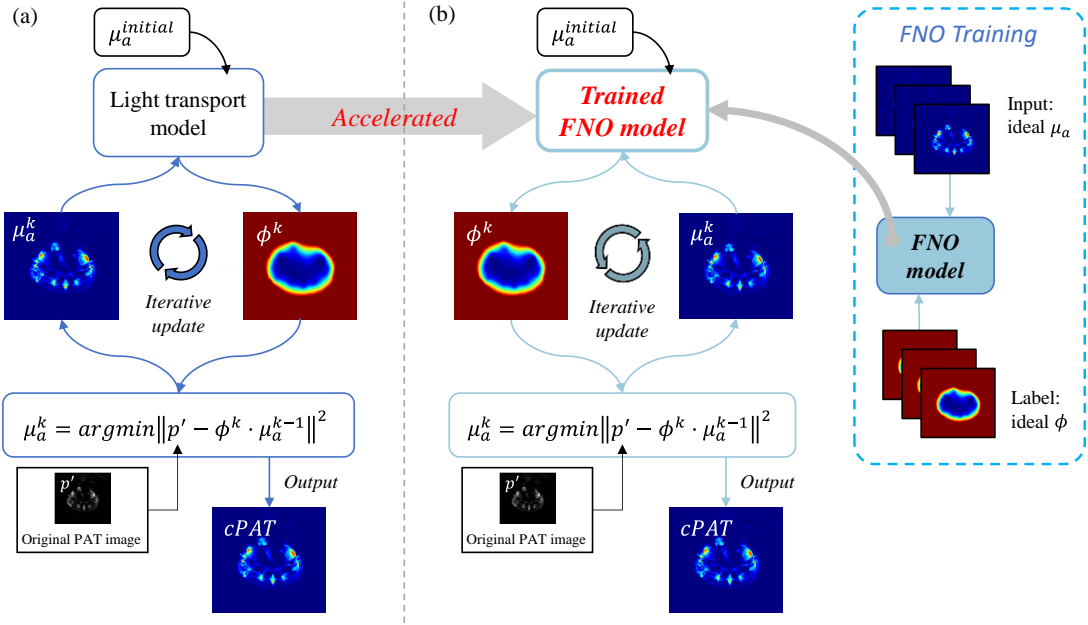

**Fig. 1.** Flow chart of iterative light fluence correction in PAT. (a) Traditional iterative LF correction method. (b) Proposed accelerated LF correction method using FNO.

### 2.3 Accelerated iterative light fluence correction by using FNO network

*2.3.1 Accelerated iterative light fluence correction*

To solve the abovementioned computational bottleneck, as shown in Fig. 1(b), we propose to use a deep-learning-based network model to learn the forward simulation capability of light transport model, and substitute the conventional numerical solvers with the trained model. Because the runtime for a well-trained network model is much faster than a numerical solver, the LF estimation time can be significantly reduced and LF correction can be further speeded up.

Fig. 2 provides a flowchart illustrating the procedure of the proposed accelerated LF correction method. We adopt the method of alternating iteration to update LF and $\mu_a$ [21]. Before the iterative correction begins, some important parameters need to be set, including $Iter1$, $Iter2$, $\varepsilon1$ and $\varepsilon2$. $Iter1$ and $Iter2$ are the iterative parameters which control the maximum number of iterations for LF map and $\mu_a$ map, respectively. $\varepsilon1$ and $\varepsilon2$ are the minimum residual errors that determine whether the iterations of LF maps and $\mu_a$ map continue, respectively. To start the iteration, an initial $\mu_a$ map is input into a trained FNO-based LF estimator to generate the corresponding LF map. Then, the conjugate gradient algorithm is used to iteratively update $\mu_a$, which is subsequently input into the FNO estimator to produce a new LF map. The above process is repeated until the loss value $Err1$, which represents the difference between the generated data and the detected data, is less than $\varepsilon1$ or when the number of LF iteration exceeds $Iter1$. At the end of the iteration process, the resulting $\mu_a$ map represents the corrected PAT image.

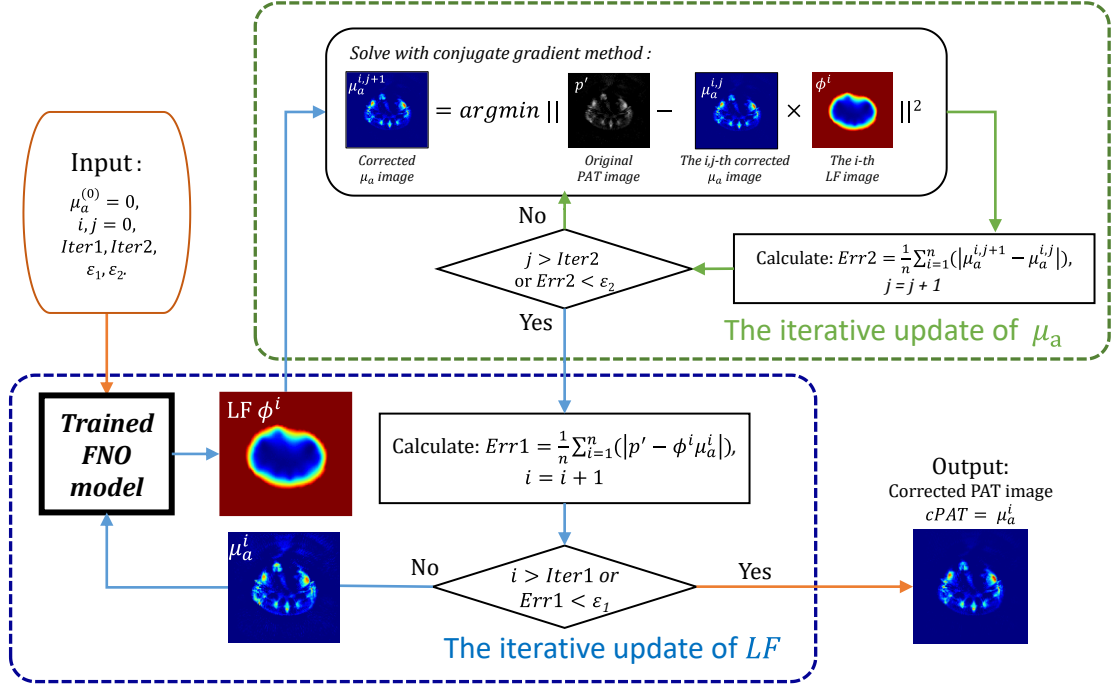

**Fig. 2.** Flow chart of accelerated iterative LF correction by FNO. $Iter1$, $Iter2$ are the max number of iterations to update LF and $\mu_a$, respectively. $\varepsilon 1$ and $\varepsilon 2$ are the minimum residual errors that control whether LF and $\mu_a$ iterations continue, respectively. Trained FNO model: a trained Fourier neural operator for LF estimation. $\phi^i$: the LF generated in the $i$-th iteration, $\mu_a^{i,j}$: the absorption coefficient generated in the $j$-th iteration of the $i$-th LF iteration

*2.3.2 Fourier neural operator for LF estimation*

At the core of our method is the Fourier neural operator, which serves as the efficient LF estimator. FNO provides a promising approach for solving complex PDE that is difficult to simulate through physical process. The architecture of the FNO network used in this study is illustrated in Fig. 3. The core component of FNO is the Fourier layer, which learns hidden information about the physical process behind the data. The input is elevated to a higher dimension determined by the hyperparameter called *channel* by a fully connected layer. Four Fourier layers iteratively update the n-*channel* underlying representation, and two convolutional layer projects the output back to the target dimension. The Fourier layer consists of two branches: the first one involves the Fourier transform, which translates feature maps into Fourier space to learn global feature information, and the other branch utilizes convolutional layers to learn high-frequency features. The first branch transforms input to the frequency domain by a two-dimensional Fourier transform and truncated by another hyperparameter, *mode*, to filter out higher-frequency components and perform linear transformations on low-frequency components. After that, two convolution layers and an activation layer are added to the first branch of the Fourier layer to further enhance low-frequency information learning. The combined output from the two branches transmits to the next layer through an activation layer. These iterative updates can be expressed as:

$$v_{t+1}(x) := \sigma(W_{t0}v_t(x) + W_{t2}(\sigma(W_{t1}\mathcal{F}^{-1}(\mathcal{F}(\kappa_\phi) \cdot \mathcal{F}(v_t))))(x)), \tag{6}$$

where $\sigma$ is a non-linear activation function whose action is defined component-wise, $W_{t0}, W_{t1}, W_{t2}$,

denotes linear transform, $\kappa_\phi$ plays the role of a kernel function which is learned from data, parameterized in Fourier space by Fourier transform $\mathcal{F}$. The features learned in the spectral space are transformed back to the spatial domain by inverse Fourier transform $\mathcal{F}^{-1}$. As the absorption coefficient varies greatly between different tissues, the $\mu_a$ map contains more high-frequency information. For that reason, we choose higher *channels* to excavate the features of $\mu_a$ maps. On the contrary, due to the slow change of diffused photon fluence in biological tissue, the LF distribution contains more low-frequency information, which informs us to choose fewer *modes* to reduce interference. In this paper, FNO (a, b) represents the FNO with a *modes* and b *channels*.

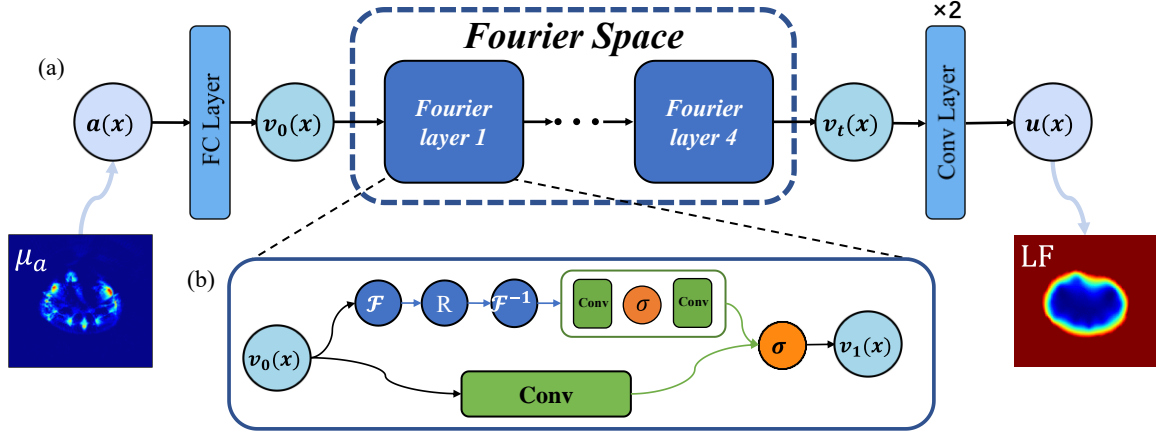

**Fig. 3.** (a) The architecture of FNO for LF estimation in PAT. The input $a(x)$ is lifted to a higher dimension using a fully connected layer and then transported into four Fourier layers. Finally, the output $v_t(x)$ is projected back to the target dimension by two 1×1 convolutional layers. (b) The detailed architecture of the Fourier layer. The high-dimension input $v_0(x)$ is transported through two different branches. The upper branch uses the Fourier transform to project the input into the Fourier domain, the linear transform R is on the low mode and filters out the high mode. Then the result projects back into the target space by inverse Fourier transform. Two 1×1 convolutional layers further enhance low-frequency information learning, and a 1×1 convolutional layer is applied in the bottom branch to perform a linear transformation of the input.

### 2.4 Implementation

The algorithm development platform utilized in this study is based on a desktop computer with a AMD Ryzen5-5600 CPU and a NVIDIA RTX 3060 GPU. The FNO model is trained using the Adam optimizer, with parameter updates based on the mean square error loss in 500 epochs. Furthermore, the learning rate is decreased by a factor of 0.1 every 25 epochs, starting from an initial value of 0.001. For all experiments, $\varepsilon 1$ and $\varepsilon 2$ are both set to $10^{-12}$.

### 3. Experimental Setup

### 3.1 PAT imaging system

The imaging equipment employed is a commercial small animal multispectral photoacoustic tomography system (MSOT inVision128, iThera Medical, Germany). Five pairs of laser emitters are evenly distributed at 270 degrees to provide 360-degree uniform illumination on the sample surface, forming a width of approximately 8 mm ring illumination, as present in Fig.4(a). The system has a tunable (660-960 nm) laser with a pulse width of around 5 ns and a repetition frequency of 10 Hz. The ultrasound generated by the excited sample is coupled through water and transmitted to a ring-shaped array transducer consisting of 128 elements covering 270 degrees with a radius of 40.5 mm. During the imaging process, the animal is placed in a specialized holder that facilitates alignment with the central axis of the ring-shaped transducer. The raw data is reconstructed into a 300×300 two-dimensional image using a model-based iterative image reconstruction algorithm [42].

**3.2 Simulation experiment**

We generate simulation datasets using simulated models of healthy mouse organs [21]. To better emulate real-world scenarios, we set the simulation conditions based on the illumination setup of the MSOT imaging system in Toast++ [40], an open-source finite-element-based LF simulation software. Specifically, as shown in Fig. 4(b), we create a circular computational mesh with a radius of 40 mm, consisting of 30 sectors and 100 rings, which resulted in 157291 nodes and 311640 elements. Five light sources were evenly distributed along a 270-degree arc with 40 mm radius. The simulated mouse is positioned at the center of the setup. The absorption coefficient and scattering coefficient of the background medium are both set to 0.0001. We obtain an uncorrected PAT image by multiplying the simulated LF map with the ideal $\mu_a$ map, and further add noise with mean 0 and variance $2\times10^{-5}$ to obtain the final uncorrected PAT image. Following this procedure, we generate five datasets at 5 different illumination wavelengths, namely, 700 nm, 730 nm, 760 nm, 800 nm and 850 nm. Each dataset contains 400 pairs of $\mu_a$ maps, LF maps, and uncorrected PAT images, of which 80% were used as training datasets, 10% as validating datasets, and 10% as test datasets. Five FNO models are trained using different datasets to learn the corresponding physical processes for different wavelengths. The ideal $\mu_a$ maps are used as input, and the ideal LF maps serve as the label for training.

To compare the performance of FNO, we train two U-Net models with the same parameters and network structure by two different methods. The first model is trained using the same training method employed for FNO and serves as an LF-estimator, which results in a U-Net-based accelerated iterative correction model, denoted as U-Net-AIC. The other model is trained to learn a direct mapping from the initial PAT image to the $\mu_a$ map, which represents an end-to-end processing by U-Net and is denoted U-Net-E2E.

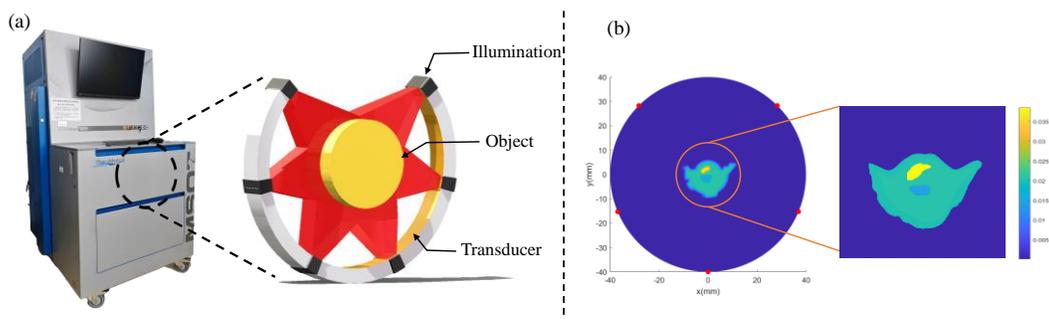

**Fig. 4.** (a) Schematic of the PAT system used in this study. (b) Left: Diagram of the simulation setup. Red dots represent the positions of the light sources. Right: absorption coefficient diagram of simulated mouse organ.

### 3.3 Animal experiment

The animal experiments are performed on the MSOT imaging system described previously. We reconstruct the raw data into PAT images using a model-based image reconstruction method [42]. As accurate $\mu_a$ and LF maps could not be obtained, we use a traditional iterative LF correction method [21] to obtain these maps simultaneously. The reason for this is that this method can obtain pixel-level reconstruction accuracy without pre-segmentation of the object. These reconstructed $\mu_a$ and LF maps are utilized as samples for training the network models and as ground truth for experimental evaluation, resulting in a total of five datasets at five different illumination wavelengths, each containing 500 PAT images, 500 $\mu_a$ maps, and 500 LF maps. As in the simulation experiments, the data sets were divided into the training sets and test sets in a ratio of 4:1. Furthermore, two U-Net models are also trained using the same method as in the simulation experiment to compare the performance of FNO.

### 3.4 Quantification metrics

Root mean square error (RMSE) and peak signal-to-noise ratio (PSNR) are used to evaluate the quality of the results produced by different methods. They are given by:

$$RMSE = \sqrt{\frac{1}{n}\sum_{i=1}^{n}(T_i - M_i)^2}, \qquad (7)$$

$$PSNR = 20\log_{10}\left(\frac{MAX}{RMSE}\right), \qquad (8)$$

where $T$ represents the ground truth, $M$ represents the result of the experiment. $MAX$ is the max value of the ground truth. All quantification metrics presented in this paper are statistical means of the results in the test set.

### 4 Results

### 4.1 Simulation results: LF estimation

Firstly, the simulation dataset with the illumination wavelength of 850 nm is used to test the forward simulation ability of our FNO-based LF estimator. Fig. 5 depicts the visual comparison of results from different networks. The absolute error maps between ground truth and the results of neural-network-based estimators are presented at the second line of Fig. 5. As can be seen, the selected networks are all capable of forward simulation, but result in different accuracy. The LF map generated by U-Net displays distinguishable artifacts, as pointed by the white arrow in the figure. In addition, the absolute error maps highlight that all results from FNO have less error compared to those from U-Net. In order to further analyze the forward simulation results of each network, we choose RMSE as the quantitative index to quantitatively compare the results of each network and listed the results at the bottom of Fig. 5, along with the number of parameters of each network. We found that the RMSE of U-Net reaches 6.14E-3, which is much higher than the RMSE of the FNO models. It is also worth noting that except for FNO (15,64), the other three FNO models with fewer parameters can achieve superior forward results than U-Net. To further assess the forward simulation capability of the proposed method, we select FNO

(10,64) as the forward network and test its performance on simulation datasets with five illumination wavelengths. The quantitative experimental results are presented in Table 1. Among all five experiments with different illumination wavelengths, the RMSE of FNO is less than that of the corresponding U-Net model. By calculation, the mean value of FNO at five illumination wavelengths is 3.79E-03, while that of U-Net is 6.18E-03, which is about 1.63 times higher than that of FNO.

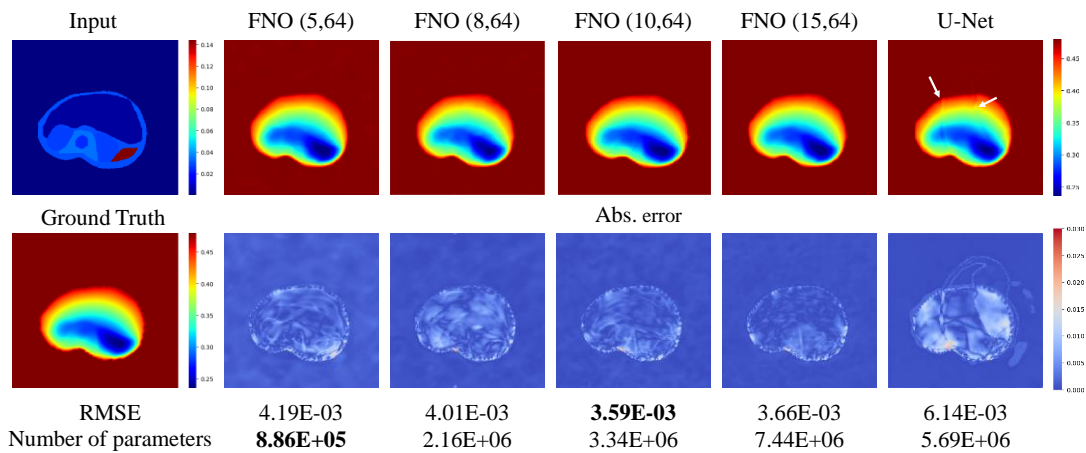

**Fig. 5.** Visual comparison of the results of FNO and U-Net for learning the forward light transport process. Input is the ideal $\mu_a$ map and Ground Truth is the ideal LF map. The bottom row shows the absolute error between the LF prediction result and ground truth. FNO (10,64) represents FNO model with 10 *modes* and 64 *channels*.

**Table 1.** RMSE between the ideal LF map and the network-generated LF map simulated by FNO and U-Net models on simulation data.

|  | 700 nm | 730 nm | 760 nm | 800 nm | 850 nm |
| --- | --- | --- | --- | --- | --- |
| FNO (10,64) | **3.59E-03** | **3.53E-03** | **3.98E-03** | **3.70E-03** | **4.15E-03** |
| U-Net | 6.14E-03 | 5.18E-03 | 6.41E-03 | 6.18E-03 | 6.98E-03 |

### 4.2 Simulation results: LF correction

Next, we compare the performance of four different LF correction methods, namely the traditional iterative correction method (TIC), U-Net-based accelerated iterative correction (U-Net-AIC), end-to-end processing by U-Net (U-Net-E2E), and our FNO-based accelerated iterative correction (FNO-AIC), in terms of correction time and correction quality. Fig. 6 shows the visual comparison of correction results at 850 nm illumination wavelength. U-Net-E2E produces the worst result, which is characterized by blurred outlines of different tissue (white arrow) and non-uniform artifacts (yellow arrow). In contrast, results obtained from iterative correction methods are much more superior and differ minimally from the ground truth. Notably, TIC produces the best results with almost negligible errors. This is because the accurate LF maps are generated by the FEM-based LF estimator. The results of the two neural-network-based iterative correction methods both show slight errors, yet FNO-AIC produces more accurate correction results compared to U-Net-AIC. This is consistent with the previous results of forward simulation capability, confirming that the error in the forward LF generation will be passed to the subsequent correction.

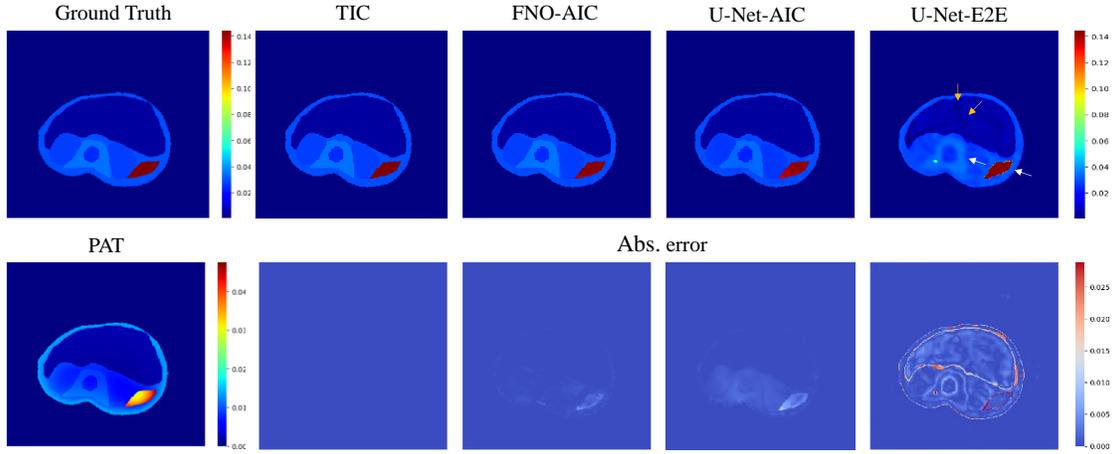

**Fig. 6.** LF Correction results of different methods in simulation experiment. Ground Truth: ideal $\mu_a$ map. PAT: uncorrected PAT image. FNO-AIC: FNO-based accelerated iterative correction method. U-Net-AIC: U-Net-based accelerated iterative correction method. U-Net-E2E: end-to-end processing method by U-Net. Abs. error: absolute error between the correction result and the Ground Truth.

Quantitative experimental results at all five illumination wavelengths are presented in Table 2. As can be seen from the table, TIC achieves the best correction effect among the four methods, but it also consumes the longest processing time, which is more than 80 seconds on average. Although U-Net-E2E is the fastest, its correction results are the worst among all methods in all experiments. The time required for the network-based accelerated iterative correction varies from 2.01 to 2.53 seconds, which is more than 30 times faster than TIC. Among the two accelerated methods, FNO-AIC achieves an average 12.08% improvement in PSNR.

**Table 2.** Quantitative results of different LF correction methods on simulation data of five different illumination wavelengths.

| Wavelength (nm) | Method | RMSE | PSNR | Time (s) |
|---|---|---|---|---|
| 700 | TIC | **0.00041** | **50.909** | 84.33 |
|  | FNO-AIC | 0.00069 | 41.667 | 2.16 |
|  | U-Net-AIC | 0.00090 | 37.936 | 2.35 |
|  | U-Net-E2E | 0.00352 | 21.955 | **0.01** |
| 730 | TIC | **0.00019** | **52.190** | 82.87 |
|  | FNO-AIC | 0.00044 | 43.535 | 2.01 |
|  | U-Net-AIC | 0.00062 | 39.213 | 2.12 |
|  | U-Net-E2E | 0.00281 | 22.690 | **0.01** |
| 760 | TIC | **0.00089** | **50.357** | 84.81 |
|  | FNO-AIC | 0.00100 | 41.213 | 2.36 |
|  | U-Net-AIC | 0.00165 | 35.994 | 2.53 |
|  | U-Net-E2E | 0.00427 | 22.671 | **0.01** |
| 800 | TIC | **0.00074** | **50.856** | 80.66 |
|  | FNO-AIC | 0.00084 | 42.310 | 2.21 |
|  | U-Net-AIC | 0.00125 | 37.106 | 2.39 |
|     | U-Net-E2E | 0.00442 | 22.694 | **0.01** |
| 850 | TIC      | **0.00138** | **50.719** | 86.06 |
|     | FNO-AIC  | 0.00193 | 39.489 | 2.31 |
|     | U-Net-AIC | 0.00208 | 35.518 | 2.50 |
|     | U-Net-E2E | 0.00521 | 22.859 | **0.01** |

**4.3 Simulation results: mesh size**

For the same imaging area, traditional LF estimators need a dense computational mesh to provide more precise LF simulation, which corresponds to more computing resources and time cost. Here, we compare the calculation speed and correction quality of the traditional method and the other three deep learning methods under three different mesh sizes. We chose the simulation dataset at 700 nm illumination wavelength as the experimental data. In the LF simulation environment, we keep the radius of the simulation environment and the number of light sources unchanged, but increase the grid parameters by 1, 1.5, and 2 times. Accordingly, the image size also increases to ensure that the size of the object remains unchanged.

The quantitative results of the above experiments are presented in Table 3. TIC yields the best correction results but consumes the longest time. Conversely, U-Net-E2E is significantly faster but produces inferior correction results. FNO-AIC demonstrates consistent precision correction capabilities for different image resolutions, whereas U-Net-AIC shows a notable increase in RMSE and a decrease in PSNR when dealing with images at 2x image size. This suggests that with the increase in image size, the processing accuracy of U-Net-AIC decreases, while that of FNO-AIC does not change. Compared to TIC, the FNO-AIC achieves acceleration rates of 39.04, 22.34, and 31.07 times under the three mesh sizes, respectively. It also outperforms the U-Net-AIC in terms of both speed and the quality of correction results. The reason for the superior performance achieved by our FNO-AIC lies in the fact that any discretized data can be transformed into the Fourier space through the application of the Fourier transform. Moreover, FNO does not require any network parameter or structure adjustments across the three image sizes, whereas U-Net necessitates corresponding modifications based on image size.

Table 3. Quantitative experimental results of different correction methods at different mesh sizes.

| | RMSE | | |
| --- | --- | --- | --- |
| | 1X | 1.5X | 2X |
| TIC | **0.00025** | **0.00031** | **0.00034** |
| FNO-AIC | 0.00050 | 0.00055 | 0.00055 |
| U-Net-AIC | 0.00064 | 0.00062 | 0.00103 |
| U-Net-E2E | 0.00316 | 0.00305 | 0.00354 |
| | **PSNR** | | |
| | 1X | 1.5X | 2X |
| TIC | **53.668** | **53.124** | **52.924** |
| FNO-AIC | 43.133 | 42.928 | 43.630 |
| U-Net-AIC | 39.279 | 39.389 | 36.224 |
| U-Net-E2E | 22.306 | 22.889 | 21.478 |
| | **Time** | | |
| | 1X | 1.5X | 2X |
| TIC | 84.33 s | 196.79 s | 447.97 s |
| FNO-AIC | 2.16 s | 8.81 s | 14.42 s |
| U-Net-AIC | 2.35 s | 10.00 s | 17.21 s |
| U-Net-E2E | **0.01 s** | **0.01 s** | **0.01 s** |

**4.4 Small animal imaging experiment result**

We conduct small animal experiments to test the effectiveness of FNO-AIC. Due to the unavailability of real $\mu_a$ and LF maps, the correction results from TIC are used as ground truth. In all small animal experiments, the FNO applied is with 10 *modes* and 64 *channels*. We set *Iter1* and *Iter2* to 30 and 20 respectively. Fig. 7 presents a visual comparison of the correction results at the liver position with two illumination wavelengths. The results from U-Net-E2E are the worst among all three methods, which show the blurred boundary of tissue (white arrow) and higher error. In contrast, the accelerated iteration methods based on neural networks generate much superior correction results. As can be seen from the error map, compared with U-Net-AIC, the error of FNO-AIC results is less at both wavelengths, especially at places with high absorption coefficient (yellow arrow), which proves that the correction result of FNO-AIC at different wavelengths is more accurate than U-Net-AIC.

Furthermore, we present the correction results of two different positions at 800 nm illumination wavelength in Fig. 8. The results of U-Net-AIC are still not ideal, which error is the largest and distributed globally. Consistent with the experimental results at different wavelengths, FNO-AIC is still more accurate than U-Net-AIC at different positions. As can be seen from the error map, the error of FNO-AIC appears less and the error value is lower compared to U-Net-AIC. In the position of high absorption coefficient (yellow arrow), FNO-AIC is also more accurate than U-Net-AIC.

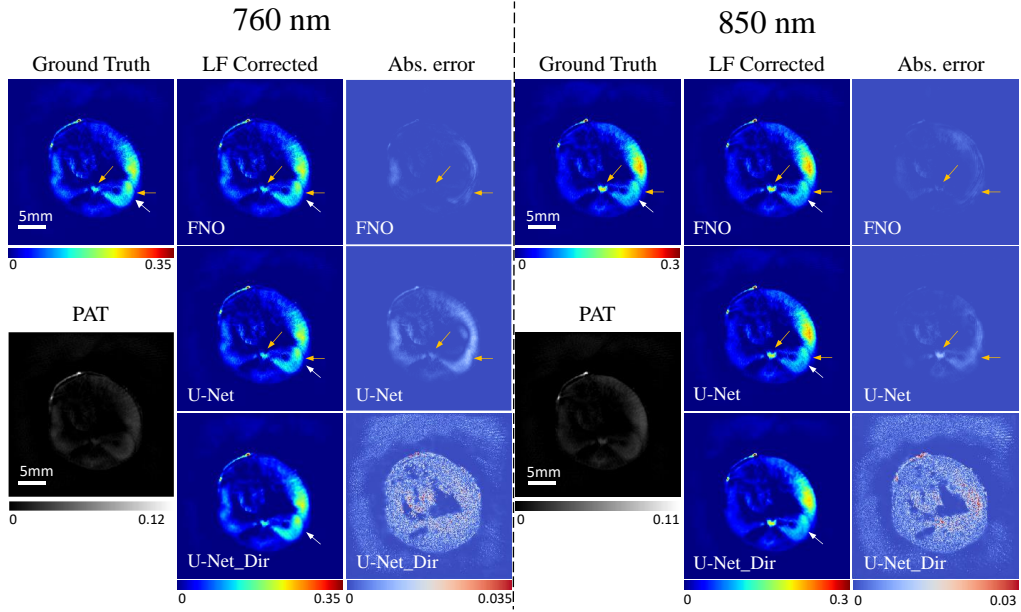

**Fig. 7.** LF correction results at the liver position with two different illumination wavelengths 760 nm and 850 nm. Ground Truth: correction results of traditional iterative correction method. PAT: uncorrected PAT image. FNO-AIC: FNO-based accelerated iterative correction method. U-Net-AIC: U-Net-based accelerated iterative correction method. U-Net-E2E: end-to-end processing method by U-Net. Abs. error: absolute error between the correction result and Ground Truth.

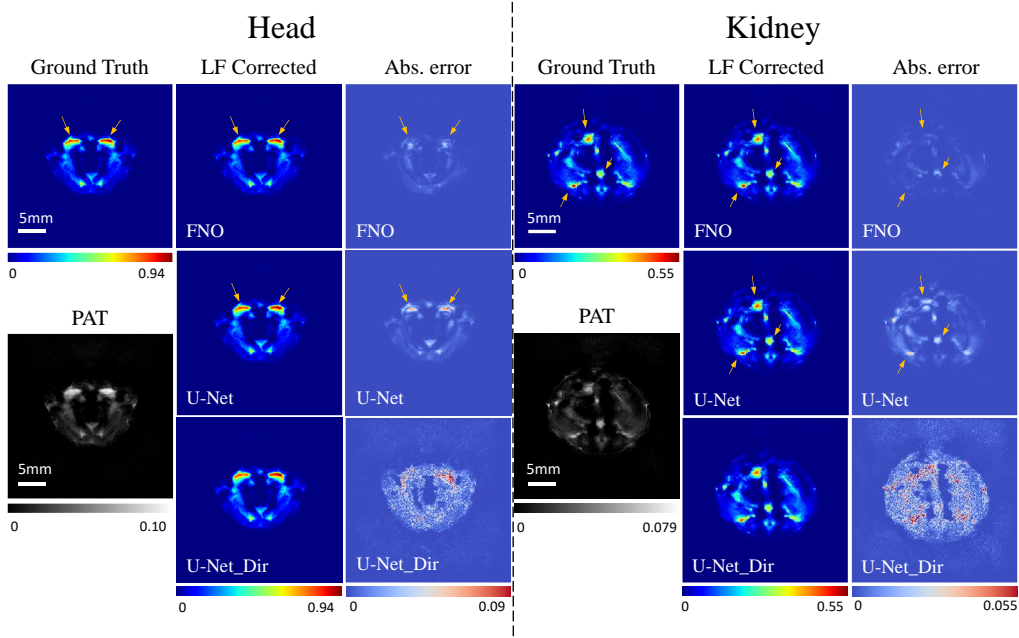

**Fig. 8**. LF correction results at two different positions. Ground Truth: correction results of traditional iterative correction method. PAT: uncorrected PAT image. Abs. error: absolute error between the correction result and Ground Truth.

The evaluation metrics of three different correction methods are quantitatively shown in Table 4. At five illumination wavelengths, FNO-AIC reaches a mean RMSE of 0.0012 and a mean PSNR of 55.274. U-Net-AIC reaches a mean RMSE of 0.0015 and a mean PSNR of 53.223. U-Net-E2E reaches a mean RMSE of 0.0077 and a mean PSNR of 34.957. The quantization results are consistent with the visual comparison results above which indicates that U-Net-E2E produces the worst correction quality at five illumination wavelengths. Compared with U-Net-AIC, the results of FNO-AIC are better under the two quantitative indicators of RMSE and PSNR at five illumination wavelengths, which proves that it has a correction ability closer to the traditional iterative method.

Table 5 shows the time consumption of processing one PAT image by four different correction methods. The average processing time of the TIC method was 84.13s at 5 wavelengths, FNO-AIC was 0.98s, U-Net-AIC was 0.95s, and U-NET-E2E was 0.01s. The accelerated iterative method is more than 80 times faster than the traditional method.

**Table 4.** Quantified correction results of animal imaging experiments at five illumination wavelengths.

| Wavelength(nm) | Method | RMSE | PSNR |
| --- | --- | --- | --- |
| | FNO-AIC | **0.0010** | **58.423** |
| 700 | U-Net-AIC | 0.0012 | 54.525 |
| | U-Net-E2E | 0.0055 | 30.906 |
| | FNO-AIC | **0.0014** | **53.842** |
| 730 | U-Net-AIC | 0.0018 | 51.351 |
| | U-Net-E2E | 0.0134 | 34.410 |
| | FNO-AIC | **0.0016** | **54.758** |
| 760 | U-Net-AIC | 0.0020 | 52.818 |
| | U-Net-E2E | 0.0081 | 37.044 |

|  | FNO-AIC | **0.0012** | **52.847** |
|---|---|---|---|
| 800 | U-Net-AIC | 0.0015 | 52.375 |
|  | U-Net-E2E | 0.0064 | 36.311 |
|  | FNO-AIC | **0.0007** | **56.503** |
| 850 | U-Net-AIC | 0.0009 | 55.044 |
|  | U-Net-E2E | 0.0049 | 36.115 |

**Table 5.** The average LF correction time in animal imaging experiments with five illumination wavelengths.

|  | 700 nm | 730 nm | 760 nm | 800 nm | 850 nm |
|---|---|---|---|---|---|
| TIC | 77.74 s | 85.61 s | 88.77 s | 83.41 s | 85.14 s |
| FNO-AIC | 0.96 s | 1.00 s | 0.98 s | 0.97 s | 0.97 s |
| U-Net-AIC | 0.93 s | 0.99 s | 0.95 s | 0.94 s | 0.94 s |
| U-Net-E2E | 0.01 s | 0.01 s | 0.01 s | 0.01 s | 0.01 s |

## 5. Discussion

The above results of both simulation and small animal imaging experiments demonstrate that the proposed method can effectively achieve accelerated LF correction for PAT. The obtained correction results achieve comparable accuracy to traditional iterative correction method, yet the processing speed has been significantly enhanced, with at least a 30-fold improvement. Notably, the proposed method demonstrates its suitability for diverse PAT images of varying sizes, without substantial compromise to processing speed. Also, the method proposed is suitable for PAT images of different illumination wavelengths and achieves consistent LF correction results.

As can be found in both simulation and small animal imaging experiments, DL-based iterative correction methods outperform end-to-end methods. This is because the network model in iterative correction learns the mapping between tissue absorption coefficient and LF, which is determined by physical light transport theory, whereas end-to-end models learn the mapping between initial pressure and absorption coefficient, which is more complicated since it is equivalent to solve two unknown independent variables by one dependent variable. Therefore, the size of neural network in our approach is smaller yet the LF correction accuracy is better.

FNO has the advantage of invariance to discretization. Therefore, compared with the traditional U-Net model that needs to adjust network parameters or structure for different input image sizes, FNO can generate results with the same quality across different image resolutions without any adjustment. Moreover, traditional FEM-based LF solver needs a larger mesh to perform more accurate simulations, which means more iteration and computation. In contrast, due to its discrete invariance, FNO can solve for the LF map under different levels of discretization without changing parameters. At the same time, its calculation speed and solution accuracy will not decrease.

Despite all these superiorities, the proposed work also has limitations and can be further improved in future works. First of all, there is still a gap between the accuracy of FNO and traditional LF solver. This is because the model has not yet fully learned the ideal characteristic of nonlinear light propagation in tissue. This limitation is expected to be improved by further modifying the network structure using recently proposed advanced techniques [45][46]. Secondly, when the light source is determined, the LF in biological tissues is determined by the absorption coefficient and the scattering coefficient together. In this paper, the relationship between the absorption coefficient and the LF is studied by default when the scattering coefficient is known. However, in practice, the scattering coefficient is unknown, and thus how to solve the absorption coefficient and scattering coefficient at the same time needs further research.

## 6. Conclusion

In this study, we propose an iterative PAT light fluence correction method accelerated by the Fourier neural operator. By substituting the conventional light transport model with a well-trained Fourier neural operator and employing an alternating iterative LF correction algorithm, the proposed method achieves precise reconstruction of the absorption coefficient map with exceptional computational efficiency. We tested the proposed method on simulation and small animal imaging experiments, and the results demonstrate the feasibility and adaptability of the proposed method.

**Funding**
National Natural Science Foundation of China (62371220); Guangdong Basic and Applied Basic Research Foundation (2021A1515012542, 2022A1515011748); Guangdong Pearl River Talented Young Scholar Program (2017GC010282).